\documentclass[preprint,prb,showpacs,preprintnumbers,amsmath,amssymb]{revtex4}

\usepackage{graphicx}
\usepackage{dcolumn}
\usepackage{bm}

\begin{document}


\title{Anomalous precursor diamagnetism at low reduced magnetic fields and the role of $T_C$ inhomogeneities in the superconductors $\rm{Pb_{55}In_{45}}$ and underdoped $\rm{La_{1.9}Sr_{0.1}CuO_4}$}

\author{Luc\'ia Cabo}
  \email{fmlucia@usc.es}
\author{F\'elix Soto} 
\author{Mauricio Ruibal}
\author{Jes\'us Mosqueira}
\author{F\'elix Vidal}

\affiliation{LBTS, Departamento de F\'isica da Materia Condensada, Universidade de Santiago de Compostela, E-15782 Santiago de Compostela, Spain}
\date{\today}
\begin{abstract}
The magnetic field dependence of the magnetization was measured above the superconducting transition in a high-$T_C$ underdoped cuprate (La$_{1.9}$Sr$_{0.1}$CuO$_4$) and in a low-$T_C$ alloy (Pb$_{55}$In$_{45}$).
Near the superconducting transition [typically for $(T-T_C)/T_C\stackrel{<}{_\sim}5\times10^{-2}$] and under low applied magnetic field amplitudes [typically for $H/H_{C2}(0)\stackrel{<}{_\sim}10^{-2}$, where $H_{C2}(0)$ is the corresponding upper critical field extrapolated to $T=0$ K] the magnetization of both samples presents a diamagnetic contribution much larger than the one predicted by the Gaussian Ginzburg-Landau (GGL) approach for superconducting fluctuations. 
These anomalies have been already observed in cuprate compounds by various groups and attributed to intrinsic effects associated with the nature of these high-$T_C$ superconductors itself. 
However, we will see here that our results in both high- and low-$T_C$ superconductors may be explained quantitatively, and consistently with the GGL behavior observed at higher fields, by just taking into account the presence in the samples of a uniform distribution of $T_C$ inhomogeneities. 
These $T_C$ inhomogeneities, which may be in turn associated with chemical inhomogeneities, were estimated from independent measurements of the temperature dependence of the field-cooled magnetic susceptibility under low applied magnetic fields. 
These conclusions are further confirmed by some additional low field magnetization measurements in a Pb$_{92}$In$_{8}$ alloy and in pure Pb.
%
%
The results summarized here also fully confirm the intrinsic character of our previous measurements of the fluctuation induced diamagnetism in La$_{1.9}$Sr$_{0.1}$CuO$_4$ under moderate magnetic field amplitudes (up to $H/H_{C2}\sim0.2$) as well as the corresponding suggestions that the precursor Cooper pairs are not affected by the presence of a pseudogap in the normal state in this underdoped cuprate.

\end{abstract}

\pacs{74.40.+k, 74.25.Ha, 74.72.Dn, 74.81.-g}
\maketitle

\section{Introduction}

The normal state magnetization near any superconducting transition will decrease due to the presence of fluctuating Cooper pairs created by the unavoidable thermal agitation energy.\cite{tinkham9} 
This effect, called precursor (or fluctuation induced) diamagnetism, was predicted by Schmidt\cite{Schmidt} and Schmid\cite{Schmid} and first observed by Tinkham and co-workers in low critical temperature, $T_C$, superconductors.\cite{tinkham9,Gollub} Since then, the precursor diamagnetism above $T_C$ was measured in different low- and high-$T_C$ superconductors.\cite{tinkham9,Skocpol,prl_pbin,gap_sym,Johnston,prl_pc_lasco,epl_ybco} 
Not too close to $T_C$, above the so-called Levanyuk-Ginzburg reduced temperature, and under low or moderate magnetic field amplitudes [for $H$ well below $H_{C2}(0)$, the upper critical field extrapolated at $T=0\: \rm{K}$], these different measurements have been explained at a quantitative level in terms of different versions (adapted to the material structure and its spatial dimensionality) of the mean field Ginzburg-Landau approach with Gaussian fluctuations of the superconducting order parameter (GGL approach).\cite{tinkham9,Gollub,Skocpol,prl_pbin,gap_sym,Johnston}
More recently, the agreement between the GGL-like approaches and the experimental results in low- and high-$T_C$ superconductors have been extended to the high reduced temperature region, above $t\equiv T/T_C\approx1.1$, by empirically introducing a so-called total energy cutoff, which takes into account the limits imposed by the uncertainty principle to the shrinkage of the superconducting wave function when the reduced temperature or the reduced applied magnetic field increases.\cite{uncertainty} 
Although the GGL approach does not formally apply  at high reduced temperatures or magnetic fields,\cite{tinkham9} this quantum constraint is expected to  be very general and it leads to  the absence of fluctuating Cooper pairs, and then of the corresponding fluctuation diamagnetism, above $t\approx1.7$, but also for reduced magnetic fields, $h\equiv H/H_{C2}(0)$, above $h\approx1.1$.
These last predictions were confirmed by measurements of the precursor diamagnetism and of the paraconductivity in different low- and high-$T_C$ superconductors, including underdoped cuprates.\cite{prl_pbin,prl_pc_lasco,epl_ybco,uncertainty,prb_pbin,Curras,Naqib} 

Although for high reduced fields and temperatures the extended GGL approach summarized above is still an open issue, the conventional approach is widely believed to be applicable under low reduced fields and temperatures, and in absence of nonlocal effects.\cite{tinkham9,Schmidt,Schmid,Gollub,Skocpol,prl_pbin,gap_sym,Johnston}   
However, recent measurements of the precursor diamagnetism performed under very low reduced magnetic fields  ($h<10^{-2}$) in various high temperature cuprate superconductors, very in particular in some underdoped materials,  seem to be beyond that conventional scenario:\cite{prb_carretta,lascialfari02,lascialfari03,Ong1} the measured  amplitude is orders of magnitude bigger than the one predicted by the GGL-like approaches and also, in some cases, these effects are appreciable at reduced temperatures well above $t\approx1.7$. 
Moreover, in some cases the magnetization versus magnetic field presents an upturn at very low fields. 
Enhanced precursor diamagnetism has been also observed under high fields, which is being related to the large Nernst signal observed above $T_C$ in high-$T_C$ superconductors.\cite{Ong2,Ong3}
These results seem to extend to the precursor diamagnetism the magnetization anomalies earlier observed below $T_C(H)$ in various high-$T_C$ cuprate superconductors\cite{Bergemann} and they could also be related to other anomalous phenomena around $T_C$ in cuprate superconductor films, as the giant proximity effect or the diamagnetic domains observed well above $T_C$.\cite{Decca, iguchi}
The interest of these striking experimental results on the precursor diamagnetism is also enhanced by the fact that they are being related to two of the at present most debated issues (entangled, in some scenarios) of the high temperature cuprate superconductors: The existence of intrinsic electronic inhomogeneities at different length scales\cite{Dagotto} and the existence of a wide temperature region above the measured $T_C$ (up to the pseudogap temperature, in underdoped materials) where the long-range phase order will be destroyed by phase fluctuations.\cite{Orenstein} In fact, the anomalies observed in the precursor diamagnetism in cuprates are being proposed as experimental evidences of the relevance of these issues in cuprate superconductors.\cite{prb_carretta,lascialfari02,lascialfari03,Ong1,Ong2,Ong3,Ovchinnikov,Sewer,demello,Anderson}

Another possible cause of at least some of the magnetization anomalies measured and studied theoretically in the works summarized above can be, however, the presence of conventional $T_C$ inhomogeneities, just associated with chemical and structural inhomogeneities. In fact, the complex chemistry of the high-$T_C$ cuprate superconductors makes that even the best available single crystals are usually not free of inhomogeneities at different length scales and with different spatial distributions.\cite{inhomo} In addition, some of the anomalies observed in other properties around $T_C(H)$ in cuprate superconductors, very in particular in the magnetoresistivity and in the thermopower, initially attributed to different intrinsic effects, were later successfully explained in terms of $T_C$ inhomogeneities associated with chemical inhomogeneities at long length scales [much bigger than the superconducting coherence length amplitude, $\xi(0)$].\cite{review_vidal} 
Moreover, in both the low-$T_C$ alloys and high-$T_C$ cuprate superconductors, the nonstoichiometric nature of the samples could lead to the presence of unavoidable chemical, \textit{intrinsiclike}, inhomogeneities.\cite{McElroy}

To provide a first example of the possible interplay between the anomalous precursor diamagnetism at low field amplitudes in high-$T_C$ superconductors and  $T_C$ inhomogeneities at long length scales associated with chemical inhomogeneities, in this paper we are going to  study these anomalies in an underdoped cuprate, the $\rm{La_{1.9}Sr_{0.1}CuO_4}$ (hereafter called LSCO). For that, we will first present detailed measurements of the magnetization  versus the applied magnetic field curves for temperatures near the transition. At low reduced temperatures, typically below $t-1\approx5\times10^{-2}$, these $M(H)_T$ curves show an anomalous behavior when the reduced field becomes below typically $h\approx10^{-2}$: at these reduced fields the $M(H)_T$ curves show an upturn and the corresponding precursor diamagnetism amplitude takes values orders of magnitude larger than the one predicted by the GGL approach. These anomalies, which have been already reported for the same compound by Lascialfari and co workers,\cite{lascialfari03} disappear at higher reduced fields and/or temperatures, where the data become in excellent agreement with our previous measurements and with the extended GGL approach.\cite{prl_pc_lasco} We will complement these measurements by checking if similar magnetization anomalies appear also in conventional (singlet s-wave pairing) BCS low-$T_C$ superconducting alloys when the samples also have a wide transition due to the unambiguous presence of chemical inhomogeneities. 
For that, we will present here measurements of the precursor diamagnetism in a somewhat inhomogeneous $\rm{Pb_{55}In_{45}}$ alloy, which demonstrate the presence of similar anomalies for equivalent reduced magnetic fields and temperatures. We also show that these anomalies are significantly reduced in a more homogeneous Pb$_{92}$In$_{8}$ alloy, and absent in pure Pb. 

In the second part of this paper, we will analyze these magnetization anomalies in the LSCO and the Pb-In superconductors in terms of $T_C$ inhomogeneities by using a simple model similar to the one proposed a long time ago by Maza and Vidal\cite{maza} to study the anomalous behavior of other observables.\cite{review_vidal,maza,magnetorho1} These analyses show that it is possible to explain at a quantitative level, and consistently with the conventional GGL behavior observed at higher reduced fields and/or temperatures, the anomalous precursor diamagnetism observed at low reduced fields by just taking into account the existence of $T_C$ inhomogeneities at long length scales (that, therefore, do not affect directly the intrinsic superconducting fluctuations) and uniformly distributed in the samples. These $T_C$ inhomogeneities follow a Gaussian distribution characterized by a mean transition temperature,  $\overline T_C$, and a transition width, $\Delta T_C$, that may be estimated from independent measurements of the field-cooled magnetization under low fields.

\section{Experimental details and results}

\begin{figure}[h]
\includegraphics[scale=0.8]{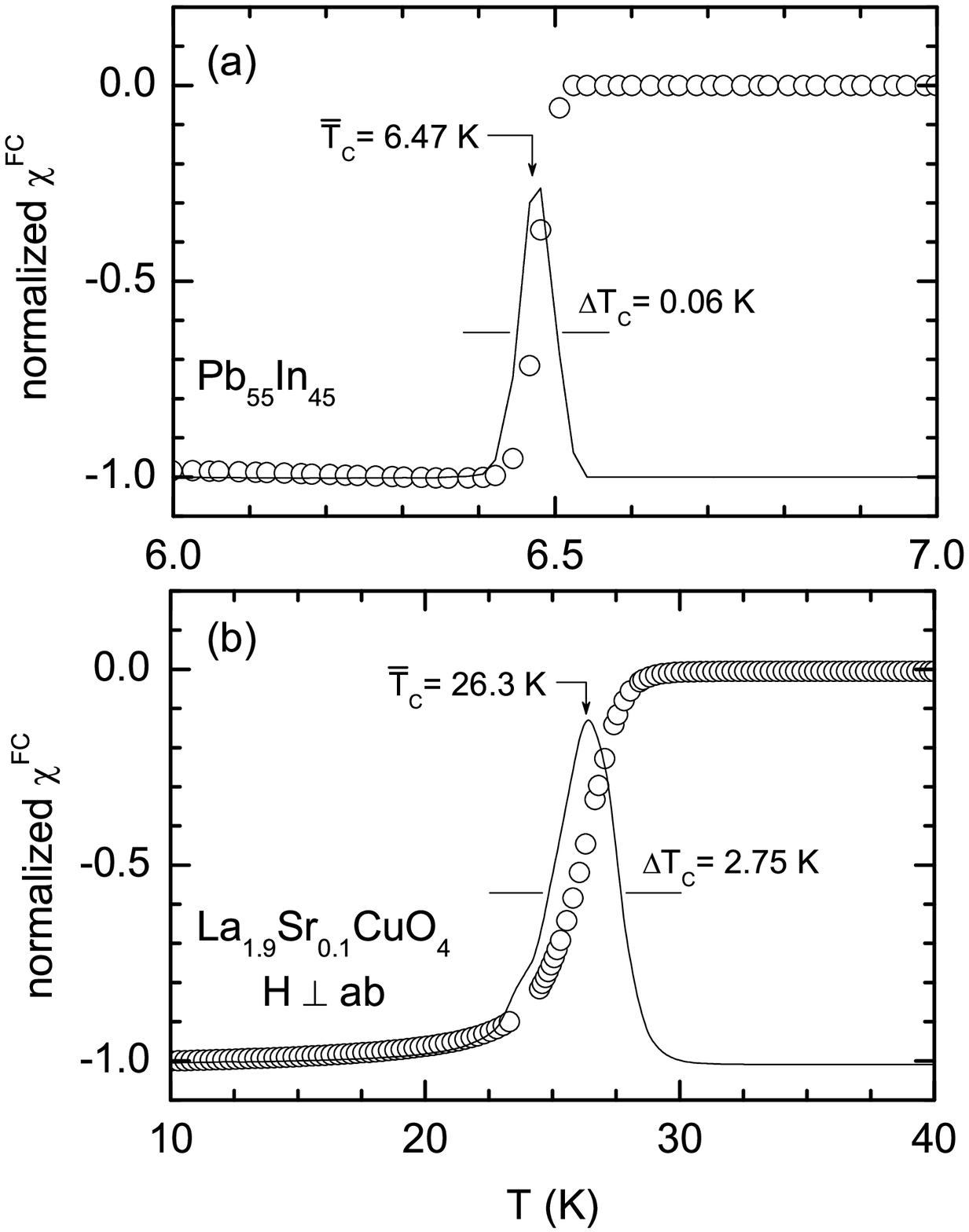}
\caption{FC magnetic susceptibility as a function of the temperature in $\rm{Pb_{55}In_{45}}$ (a) and $\rm{La_{1.9}Sr_{0.1}CuO_4}$ (b), obtained with an applied magnetic field of 0.1 mT. $\overline T_C$ and $\Delta T_C$ were estimated as the maximum position and, respectively, the full width at half maximum of the $d\chi^{FC}/dT$ curves (represented as solid lines in arbitrary units).} 
\label{FC}
\end{figure}

The LSCO sample used in the experiments was a composite made of 5-10 $\mu$m grains embedded into a low-magnetic-susceptibility epoxy (EPOTEK-301) with their $c$ crystallographic axis aligned. 
The Pb-In samples were cylinders of $\sim$5 mm in diameter and $\sim$5 mm in height. The details of their fabrication and of their general characterization may be found in Refs. \onlinecite{prl_pc_lasco}, \onlinecite{prb_pbin}, \onlinecite{crossing} and  \onlinecite{Vidal_pbin}, and in the references therein. 
Their critical temperatures were determined from measurements of the field-cooled (FC) magnetic susceptibility $\chi^{FC}$ versus temperature, obtained with an external magnetic field of $\sim$0.1 mT. 
These measurements, as well as the other magnetization measurements presented here, were performed with a SQUID magnetometer (Quantum Design, model MPMS). 
An example of the resulting $\chi^{\rm{FC}}(T)$, corresponding to the LSCO and the Pb$_{55}$In$_{45}$ samples is presented in Fig. \ref{FC}. 
These measurements are already corrected for demagnetizing effects and normalized to the ideal value of $-1$. As may be clearly seen, both superconducting transitions are somewhat broadened. 
The mean transition temperatures $\overline T_C$ were estimated from the maximum of the $d\chi^{\rm{FC}}(T)/dT$ curves (solid lines in Fig. \ref{FC}), while the transition widths $\Delta T_C$ were estimated from the corresponding full width at half-maximum. 
These $\overline T_C$ and $\Delta T_C$ values are compiled in Table I. 
The observed transition broadening is in part due to extrinsic effects like the use in the measurements of a finite magnetic field, or finite size effects (in the case of the granular LSCO sample). 
However, these effects can explain only $\sim 20\%$ of the $\Delta T_C$ value for the LSCO sample, and $\sim 30\%$ in the case of the Pb$_{55}$In$_{45}$ sample.\cite{broadening} 
Thus, we attribute the remaining part to $T_C$ inhomogeneities. 
The relative spatial variations in the Sr and In concentrations needed to justify these $\Delta T_C$ values are, respectively, $\sim3$\% and $\sim6$\%, which are very difficult to be detected by standard characterization methods. Other superconducting parameters related to the analysis of the thermal fluctuations like $H_{C2}(0)$ and the Ginzburg-Landau parameter $\kappa$, were obtained in Refs. \onlinecite{prl_pc_lasco,prb_pbin}, and \onlinecite{prb_lasco}, and are also compiled in Table I.

In Figs. \ref{pico_exp}(a) and \ref{pico_exp}(b) we present the as-measured magnetization versus applied magnetic field at a few temperatures near $\overline T_C$, for both samples studied. 
For the LSCO sample, the magnetic field was applied perpendicular to the CuO$_2$ planes.
These measurements were performed by first zero-field cooling (ZFC) the samples to the desired temperature and then taking data points for increasing magnetic fields.  
Before each $M(H)$ measurement, any environmental magnetic field or the coil's remnant magnetic field were characterized and compensated. 
For that, we took advantage of the linear $H$-dependence of $M$ in the Meissner region for temperatures well below-$T_C$ (safe from the possible effect of $T_C$ inhomogeneities).
In the insets of Fig.~2 it is presented an overview for $h$ up to $\sim 0.05$. As it is clearly seen, the magnetization is linear in a wide region. 
This behavior is compatible with the linear magnetic field dependences of the contributions to the magnetization coming from the samples' normal state, from the sample holder, from the low-susceptibility epoxy (in the case of the LSCO sample), and also from the superconducting fluctuations in the low field limit.
However, for reduced fields below $\sim 10^{-2}$ it is observed a marked deviation of the linearity in the form of a sharp diamagnetic peak. 
This anomaly is progressively reduced by increasing the temperature above $\overline T_C$ and, as commented above, is similar to the one observed in Ref. \onlinecite{lascialfari02} in oriented powders of underdoped Y$_{1-x}$Ca$_x$Ba$_2$Cu$_3$O$_y$, and in Ref. \onlinecite{lascialfari03} also in a La$_{1.9}$Sr$_{0.1}$CuO$_4$ single crystal. It is remarkable that in spite of the big differences between the LSCO and Pb-In characteristics (see Table I), this anomaly is qualitatively similar in both samples, which suggests a common origin in terms of $T_C$ inhomogeneities. 

To check this last proposal, we have performed additional measurements in a more homogeneous alloy (Pb$_{92}$In$_8$), and in pure Pb (Goodfellow, 99.9999\% purity). In Fig. \ref{plomos} we compare the low-field $M(H)$ behavior in these two last samples with the one in Pb$_{55}$In$_{45}$. These measurements were performed for the three samples 0.03$\pm0.01$ K above the corresponding $\overline T_C$. The transition widths for these three samples are shown in the inset. Although the results may be somewhat affected by the $T_C$ uncertainties, they clearly show at a qualitative level that the reduction of the anomaly amplitude is well correlated with the decrease of the transition width, i.e., with the chemical homogeneity improvement.

\section{The magnetization above the superconducting transition in presence of $T_C$ inhomogeneities}

To calculate the magnetization above the superconducting transition in presence of $T_C$-inhomogeneities at long length scales, much larger than $\xi(0)$, we will use a simple approach similar to the one already proposed by Maza and Vidal to analyze the transport properties around $T_C$ in inhomogeneous superconductors.\cite{maza} In this model the volume fraction of domains having a critical temperature $T_C$ follows a Gaussian distribution characterized by a mean transition temperature, $\overline T_C$, and a full width at half maximum, $\Delta T_C$,

\begin{equation}
\delta(T_C,\overline T_C,\Delta T_C)=\frac{\exp\left[-\left(\frac{T_C-\overline T_C}{a\Delta T_C}\right)^2\right]}{\int_0^{T_C^{\rm max}}dT_C\exp\left[-\left(\frac{T_C-\overline T_C}{a\Delta T_C}\right)^2\right]},
\end{equation}
where $a=1/2\sqrt{\ln{2}}\approx0.60$, and $T_C^{\rm max}$ is the maximum critical temperature of the superconducting system studied (7.2 K for the Pb-In alloys, and $\sim39$ K for LSCO). The measured magnetization is then given by

\begin{equation}
\left\langle M\right\rangle=\int_0^{T_C^{\rm max}} dT_C \delta(T_C,\overline T_C,\Delta T_C)M(T_C),
\label{integral}
\end{equation}
where $M(T_C)$ represents the \textit{intrinsic} magnetization of an homogeneous superconducting domain having a critical temperature $T_C$. The $H$ dependence of $T_C$ may be taken into account through $\overline T_C(H)=\overline T_{C}[1-H/H_{C2}(0)]$ where, for simplicity, we have assumed that the $H_{C2}(0)$ value is not appreciably affected by the inhomogeneities. 

An schematic representation of this simple approach may be seen in Fig. \ref{explicacion}(a). The dashed areas represent the indetermination in the lower critical field, $H_{C1}(T)$, and in $H_{C2}(T)$ due to the $T_C$ distribution [represented in Fig. \ref{explicacion}(b)]. 
When measuring the magnetization as a function of the magnetic field at temperatures slightly above $\overline T_C$, the $M(H)_T$ curves will penetrate the dashed areas at low fields and they will be then affected by a full superconducting contribution giving rise to the anomalous diamagnetic peak. Roughly, for $T\sim\overline T_C$ the diamagnetic peak will appear for a magnetic field close to $H_{C1}(0)\Delta T_C/\overline T_C$, which is $\sim0.35$ mT for the Pb$_{55}$In$_{45}$, and $\sim1.5$ mT for the LSCO sample, in agreement with the results of Figs. \ref{pico_exp}(a,b). For a more thorough comparison of the inhomogeneity model with the experimental data, the temperature and magnetic field dependence of $M(T_C)$ in Eq. \ref{integral} has to be estimated. This is the task of the two following sections. 

\subsection{Intrinsic $M(T,H,T_C)$ response of Pb$_{55}$In$_{45}$}

The intrinsic behavior of the magnetization for $T<T_C(H)$ was obtained from measurements at some constant temperatures well below $\overline T_C$ (see Fig. \ref{madres}). These measurements, like the ones in Fig. \ref{pico_exp}, were obtained under ZFC conditions and with increasing magnetic fields. As the temperatures used in these experiments are farther from $T_C$ than $\Delta T_C$, the effect of the inhomogeneities is expected to be negligible and we will consider these measurements as the {\it intrinsic} ones corresponding to a critical temperature given by $\overline T_C$. The magnetization and the magnetic field in Fig. \ref{madres} are presented in the reduced coordinates $m\equiv M/H_{C2}(0)$ and $h$, and scaled by the term $1-t$. This scaling may be justified if the GL parameter is almost temperature independent (see the next section), as it is the case for the Pb-In alloys studied here.\cite{scaling} As may be clearly seen, the measurements corresponding to different temperatures fall into a universal curve, $y=f(x)$ which, once parametrized, may be used to obtain the $m$ dependence on $t$ and $h$ in the mixed state through
\begin{equation}
m=(1-t)f\left(\frac{h}{1-t}\right).
\end{equation}

In the normal state, i.e., for $T>T_{C}(H)$ the only superconducting contribution is the one due to the thermal fluctuations. For isotropic superconductors (like the Pb-In alloys) the GGL approach in the zero magnetic field limit [i.e., for $H\ll H_{C2}(0)$] and under a total energy cutoff leads to\cite{prl_pbin,epl_ybco}
\begin{equation}
m=-\frac{\mu_0k_BT\xi(0)}{3\phi_0^2}h\left(\frac{\arctan
\sqrt{(c-t+1)/(t-1)}}{\sqrt{t-1}}
-\frac{\arctan\sqrt{(c-t+1)/c}}{\sqrt c}\right),
\end{equation}
where $\mu_0$ is the vacuum magnetic permeability, $k_B$ the Boltzmann constant, $\phi_0$ the flux quantum, and $c\approx 0.6$ is the cutoff constant.\cite{prl_pbin,cutoff} The magnetization under finite magnetic fields may be approximated by just introducing in Eq.(4) the critical temperature dependence on $H$ through the substitution $T_{C}\to T_C(H)=T_{C}[1-H/H_{C2}(0)]$, or equivalently 
\begin{equation}
t\to \frac{t}{1-h}.
\end{equation}
The magnetization given by Eqs. (4) and (5) presents an unphysical divergence at $T_C(H)$ which is a consequence of the use of the Gaussian approximation in the GL theory. This was solved by cutting off the magnetization to its value at the Levanyuk-Ginzburg temperature $t_{\rm{LG}}$ above which the Gaussian approximation is expected to be adequate.\cite{t_ginzburg} In the case of Pb-In, $t_{\rm{LG}}\approx1+10^{-4}$.

Finally, to the above superconducting contributions to the magnetization we added the normal-state or background contribution ($m_B\approx -1.28\times10^{-5}h$) which was obtained from measurements carried out at magnetic fields much larger than $H_{C2}(0)$, where the effect of the fluctuations is negligible.

\subsection{Intrinsic $M(T,H,T_C)$ response of the LSCO sample}

In the case of LSCO, the very high upper critical magnetic field does not allow a direct measurement of the $M(H)$ curve for temperatures well below $\overline T_C$. To avoid this difficulty, we have approximated it by its theoretical \textit{reversible} magnetization.\cite{reversibility} We distinguish three zones separated by the upper and lower  critical magnetic field versus temperature lines: 

	(i) In the Meissner region, i.e., for $H<H_{C1}(T)$ the magnetization may be expressed in reduced coordinates through
\begin{equation}
m=-\frac{h}{1-D},
\end{equation}
where $D$ is the demagnetizing factor. As the grains in the sample are roughly spherical, we took $D\approx 1/3$ so that $m\approx -1.5 h$.

	(ii) In the mixed state, i.e., for $H_{C1}(T)<H<H_{C2}(T)$, if $H\stackrel{>}{_\sim}0.3H_{C2}(T)$ the \textit{reversible} magnetization is given by the Abrikosov expression, that may be expressed as
\begin{equation}
m=\frac{h-1+t}{\beta_A(2\kappa^2-1)}
\end{equation}
where $\beta_A=1.16$ for a triangular vortex lattice. If $H< 0.3H_{C2}(T)$, it may be approximated by the London equation, that may be expressed as
\begin{equation}
m=-\alpha\frac{1-t}{4\kappa^2}\ln\left(\eta\frac{1-t}{h}\right). 
\end{equation}
where $\alpha\approx0.77$ and $\eta\approx1.44$.\cite{hao_clem} Due to the high anisotropy and the relatively high critical temperature of LSCO, thermal fluctuation effects are observable in the mixed state magnetization, mainly near the $H_{C2}(T)$ line.\cite{fluc_vortices} However, as an approximation, we will neglect that contribution and use Eqs. (7) and (8) as representative of the mixed state of this compound. 

In the inset of Fig. \ref{madres} it is represented the magnetic field dependence of the magnetization below $T_C$ as given by Eqs. (6-8) (in the reduced coordinates $m$ and $h$, and scaled by $1-t$). In this representation the curves corresponding to different reduced temperatures fall into a universal curve that depends only on the GL parameter (in this case we used $\kappa=60$, see Table I). 

	(iii) In the region $T>T_C(H)$, the only superconducting contribution is the one coming from the evanescent Cooper pairs created by thermal fluctuations. This contribution was again estimated through the GGL approach. In the case of highly anisotropic layered superconductors, for $H$ applied perpendicular to the superconducting CuO$_2$ (ab) planes and under a total energy cutoff, this approach leads to \cite{prl_pc_lasco}

\begin{equation}
m=-\frac{\pi\mu_0k_BT\xi_{ab}^2(0)}{3\phi_0^2s}h\left(\frac{1}{t-1}-\frac{1}{c}\right),
\end{equation}
where $s=0.66$ nm is the ab planes periodicity length, $\xi_{ab}(0)=3.0$ nm the in-plane GL coherence length amplitude, and $c\approx0.6$ is the cutoff constant.\cite{cutoff} This equation is valid only in the zero-field limit [$H\ll H_{C2}(0)$], but again, its validity may be crudely extended to finite magnetic fields by just taking into account the $T_C$ dependence on $H$ by replacing $t$ by $t/(1-h)$. The divergence at $T_C(H)$ was also avoided by cutting off the magnetization to its value at the Levanyuk-Ginzburg temperature, which for LSCO is around $t_{\rm{LG}}\approx1+3\times10^{-2}$.\cite{t_ginzburg} Finally, the normal-state or background contribution ($m_B\approx -6.4\times10^{-5}h$) was obtained from measurements carried out at temperatures well above $T_C$, where the effect of the fluctuations is negligible.

\section{Data analysis}

\begin{figure*}[h]
\includegraphics[scale=1]{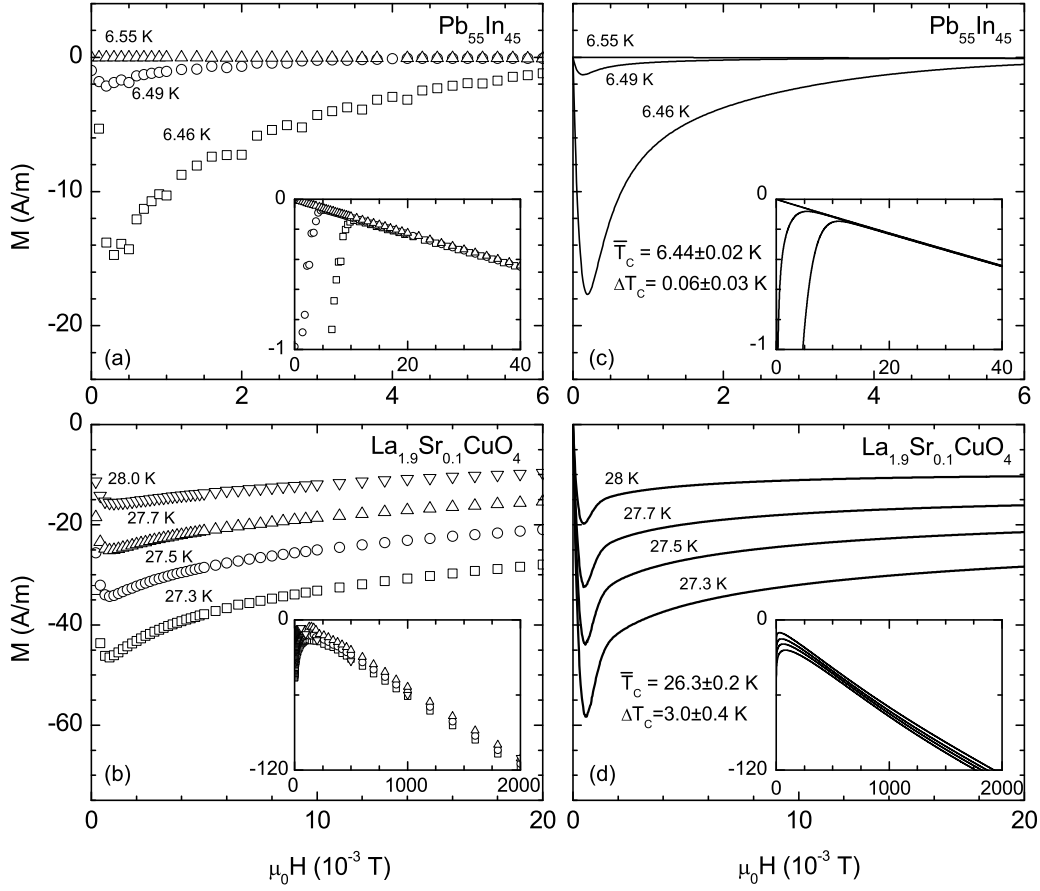}
\caption{Magnetic field dependence of the as-measured magnetization for different temperatures near $\overline T_C$ in $\rm{Pb_{55}In_{45}}$ (a) and $\rm{La_{1.9}Sr_{0.1}CuO_4}$ (b). The diamagnetic anomaly is clearly seen at very low magnetic field amplitudes. In the overview presented in the insets it is clearly seen that at higher magnetic fields the magnetization recovers the expected linear behavior (see the main text for details). For clarity, the curves fitted to these data points from the inhomogeneities model [Eq.(2)] are shown in separated figures [(c) and (d)]. The values obtained for the fitting free parameters ($\overline T_C$ and $\Delta T_C$, indicated in the figure) are in good agreement with those extracted from the low-field $\chi^{FC}(T)$ measurements presented in Fig. \ref{FC} (see Table I).}
\label{pico_exp}
\end{figure*}

\begin{figure}[h]
\includegraphics[scale=0.8]{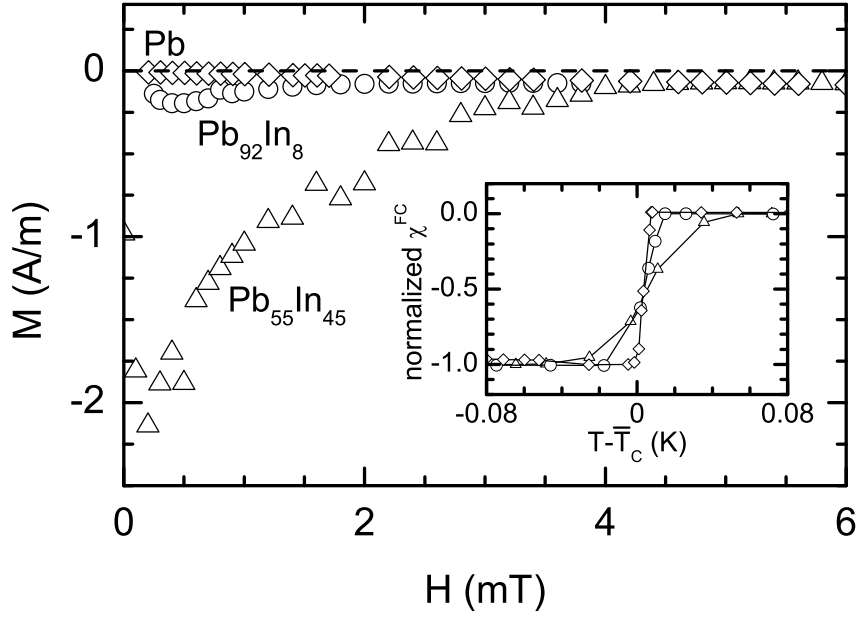}
\caption{Magnetic field dependence in the very low field range of the magnetization in Pb$_{55}$In$_{45}$, Pb$_{92}$In$_8$, and in pure Pb. The three curves were measured $0.03\pm0.01$ K above the corresponding $\overline T_C$. The 
respective $\overline T_C$ and $\Delta T_C$ are summarized in Table I. The alloys present 
a diamagnetic anomaly, which is absent in pure Pb. In the inset, 
it is represented the normalized $\chi^{FC}$ as a function of $T-\overline T_C$, to show the 
broadening of the transition in each sample.}
\label{plomos}
\end{figure}

\begin{figure}[h]
\includegraphics[scale=0.8]{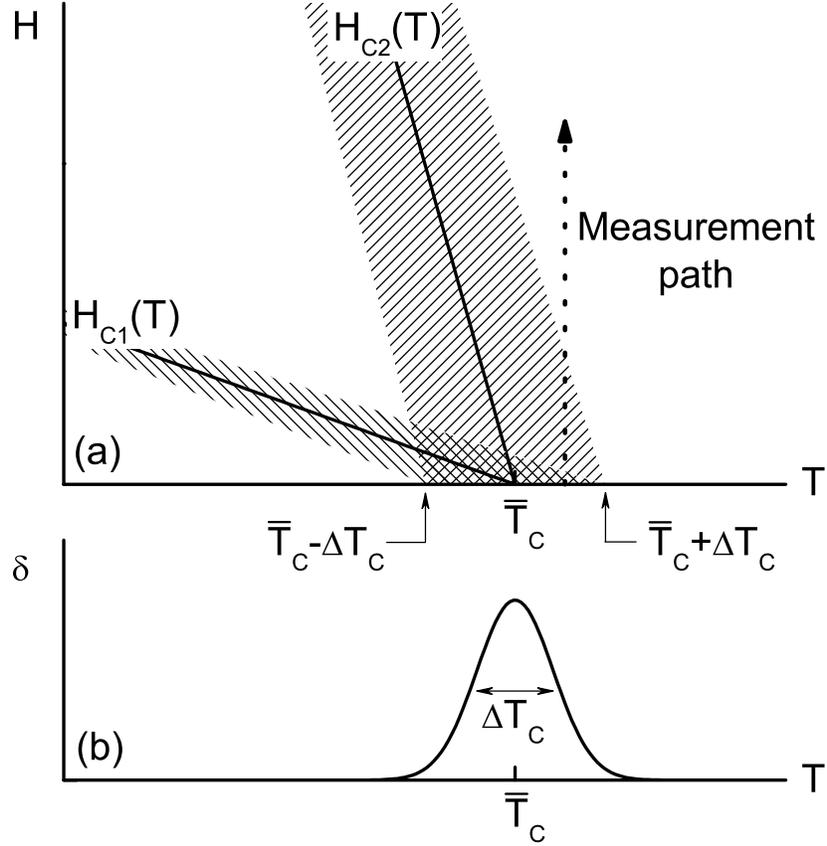}
\caption{(a) Schematic phase diagram for an inhomogeneous sample with a $T_C$ distribution like the one shown in (b). The dashed areas represent the broadening of the $H_{C1}(T)$ and $H_{C2}(T)$ lines due to the $T_C$ distribution. At low field amplitudes, and near $T_C$, the $M(H)_T$ curves (like the ones in Fig. \ref{pico_exp} (a) and (b)) will be affected by a mixed state contribution in the dashed area, and even by a full diamagnetic contribution in the crossed area.} 
\label{explicacion}
\end{figure}

\begin{figure}[h]
\includegraphics[scale=0.8]{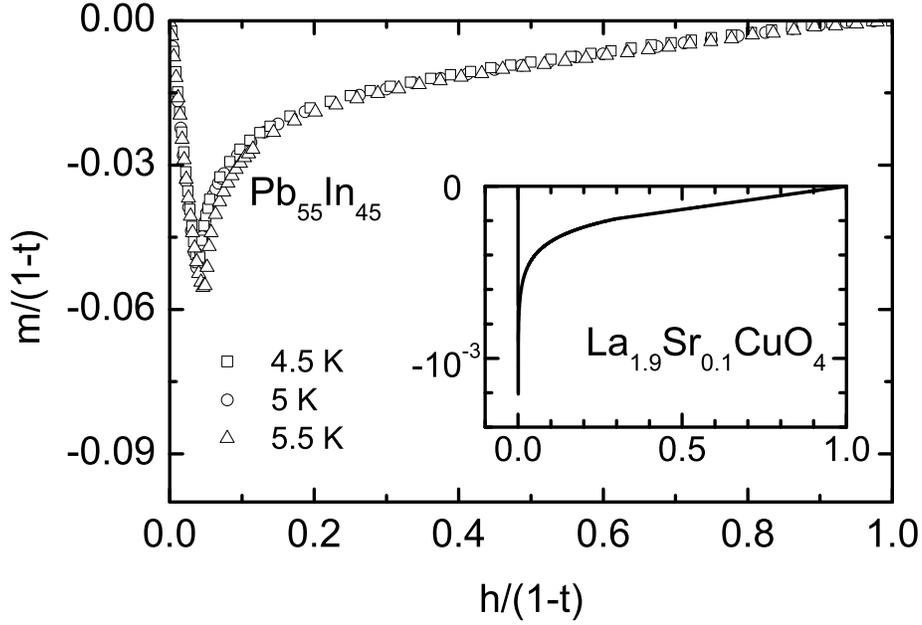}
\caption{Magnetic field dependence of the magnetization of Pb$_{55}$In$_{45}$ (in reduced coordinates) for temperatures well below $\overline T_C$. These measurements were obtained under ZFC conditions and with increasing magnetic fields. The use of these reduced coordinates collapse the curves corresponding to different temperatures (see the main text for details). In the inset, the curve for $\rm{La_{1.9}Sr_{0.1}CuO_4}$ was calculated from Eqs. (6) to (8) by using $\kappa=60$.}
\label{madres} 
\end{figure}

In Figs. \ref{pico_exp}(c) and \ref{pico_exp}(d) we present the results of our model of $T_C$ inhomogeneities for the magnetic field dependence of the magnetization near the average critical temperature. 
These curves were obtained by fitting Eq.(2) to the data of Figs. \ref{pico_exp}(a) and \ref{pico_exp}(b), by using the intrinsic magnetization curves for LSCO and Pb-In obtained in the precedent section. 
The only free parameters were the mean critical temperature $\overline T_C$ and the transition width $\Delta T_C$. As may be clearly seen, in spite of its simplicity, this model reproduces quantitatively and consistently the low-field diamagnetic anomaly observed in both samples in isotherms close to $\overline T_C$. 
It also reproduces its reduction (and even its disappearance) at higher temperatures. 
The resulting $\overline T_C$ and $\Delta T_C$ values (indicated in the figures) are in good agreement with the low-field $\chi^{\rm{FC}}(T)$ measurements presented in Fig. \ref{FC}. 
The root mean square, rms, relative deviation with respect to the experimental data points are $\sim30$\% in the case of the Pb-In alloy, and $\sim15$\% for the LSCO sample. 
In the case of LSCO, the use of a different intrinsic magnetization curve within the uncertainty in $\kappa$ ($\sim20$\%) or in $H_{C1}$ ($\sim30$\%) could change the resulting $\overline T_C$ and $\Delta T_C$ values within $\pm3$\% and $\pm10$\%, respectively. 
However, the \textit{rms} relative deviation of the $T_C$-inhomogeneities model with respect to the experimental data would remain below $\sim20$\%.

Let us remark that the temperature dependence of the magnetic field at which the diamagnetic peak occurs, $H_{\rm peak}$, is very well described by just taking into account the $T_C$ inhomogeneities. 
This conclusion probably applies also to the previous measurements of Lascialfari \textit{et al.} (Ref. \onlinecite{lascialfari03}) in a La$_{1.9}$Sr$_{0.1}$CuO$_4$ single crystal.\cite{Lascialfari_comment}
Other features of the diamagnetic anomalies, like the irreversibility observed in the $M(H)_T$ curves for magnetic fields near $H_{\rm peak}$,\cite{lascialfari03} may also be accounted for by the presence of $T_C$ inhomogeneities by just taking into account the irreversibility of the mixed state magnetization.

\begin{figure}[hb]
\includegraphics[scale=0.8]{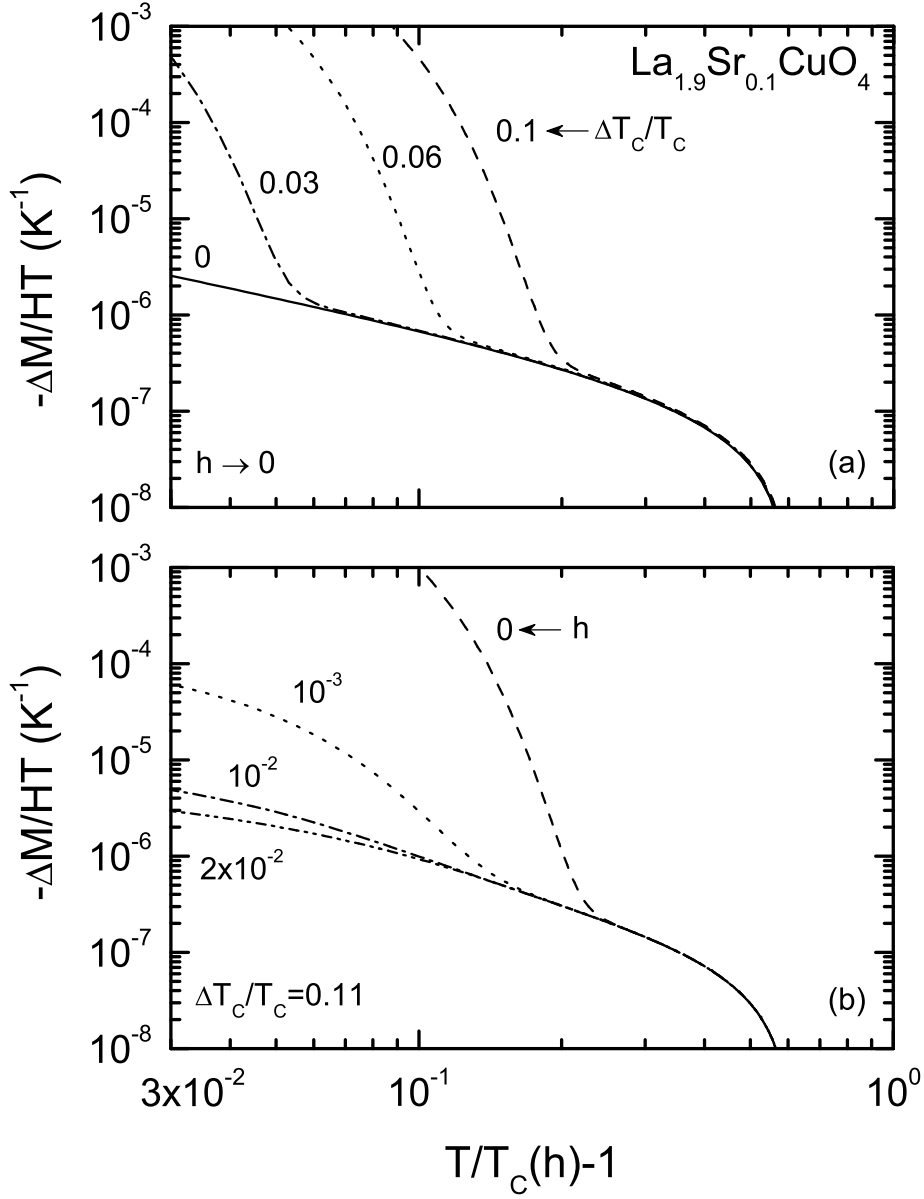}
\caption{Example, corresponding to $\rm{La_{1.9}Sr_{0.1}CuO_4}$, of the reduced temperature dependence of the fluctuation-induced magnetic susceptibility (over $T$) above $\overline T_C$. These curves were calculated by using the $T_C$ inhomogeneities model in different situations: The curves in (a) correspond to the $h\to0$ limit and different $\Delta T_C/\overline T_C$ values. The ones in (b) were obtained by using the $\Delta T_C/\overline T_C$ value corresponding to our $\rm{La_{1.9}Sr_{0.1}CuO_4}$ sample and different $h$ values. As may be clearly seen, an applied magnetic field as low as $h\approx2\times10^{-2}$ quenches the effect of the $T_C$ inhomogeneities even for temperatures very close to $\overline T_C(h)$ (see the main text for details).} 
\label{deltachi}
\end{figure}

Although our present paper is centered on the magnetic field behavior of $M$, the influence of the $T_C$ inhomogeneities on the reduced temperature dependence of the fluctuation induced magnetization $\Delta M$ is illustrated in Fig. \ref{deltachi}. In the upper panel we present an example of $\Delta M/H$ [for convenience over $T$, see Eq. (9)] against $\overline t-1\equiv T/\overline T_C-1$, for different $\Delta T_C/\overline T_C$ values. These curves were calculated by using the $T_C$-inhomogeneities model [Eq.(2)] with the superconducting parameters adequate for LSCO, and in the $h\to 0$ limit. As expected, the reduced temperature above which the effect of the $T_C$ inhomogeneities is negligible is given by $\overline t-1\sim2\Delta T_C/\overline T_C$ (i.e., by $T\approx\overline T_C+2\Delta T_C$). By using $\Delta T_C/\overline T_C\approx0.11$, as corresponds to our LSCO sample, the experimental window for the analysis of the fluctuation effects when $h\to0$ would be reduced to $\overline t-1\stackrel{>}{_\sim}0.22$. 
This situation changes drastically in presence of a finite magnetic field, as shown in Fig. \ref{deltachi}(b), where $-\Delta M/HT$ is plotted against $\overline t(h)-1$ for different $h$-values and for $\Delta T_C/\overline T_C=0.11$. The use in this figure of an $h$-dependent reduced temperature compensates the $\overline T_C$ shift due to the applied magnetic field, and the differences between the different curves are just due to the $T_C$ inhomogeneities. As may be clearly seen, an increasing magnetic field reduces progressively the effect of the $T_C$ inhomogeneities so that for $h$ as low as $\sim2\times10^{-2}$ the differences with respect to the ``homogeneous'' curve fall below 30\% in all the accesible reduced temperature region. This result validates the earlier analyses on the fluctuation effects in LSCO for finite applied magnetic fields presented in Ref. \onlinecite{prl_pc_lasco}. Also, it illustrates the difficulties that may arise when studying fluctuation effects at low reduced temperatures and, simultaneously, low reduced magnetic fields.

\section{Conclusions}

We have presented measurements of the magnetization versus magnetic field for temperatures just above the superconducting transition in a high-$T_C$ underdoped cuprate (La$_{1.9}$Sr$_{0.1}$CuO$_4$) and in a low-$T_C$ alloy (Pb$_{55}$In$_{45}$). Near the superconducting transition (for $t-1\stackrel{<}{_\sim}10^{-2}$) and under low field amplitudes, i.e., for $h\stackrel{<}{_\sim}10^{-2}$, the $M(H)_T$ curves present an anomalous diamagnetic contribution that cannot be explained in terms of the GGL approach for thermal fluctuations in homogeneous superconductors. This anomaly disappears at higher reduced  magnetic fields, and the fluctuation induced magnetization turns out to be in excellent agreement with the GGL approach. Enhanced diamagnetism has been previously observed by other groups in different underdoped HTSC.\cite{prb_carretta,lascialfari02,lascialfari03,Ong1} Our measurements were quantitatively and consistently explained by a simple model of $T_C$ inhomogeneities at length scales much bigger than $\xi(0)$ and uniformly distributed. These inhomogeneities are expected to come from spatial variations in the concentration of La and Sr in the LSCO sample, and of the two constituents in the Pb-In alloys. In our model, the inhomogeneities follow a Gaussian distribution characterized by a mean transition temperature $\overline T_C$ and a transition width $\Delta T_C$. We found that the $\Delta T_C$ and $\overline T_C$ values needed to explain the observed anomalies are in excellent agreement with the ones obtained from the temperature dependence of the field cooled magnetic susceptibility around $T_C$ under low applied magnetic fields. The associated \textit{local} variations in the Sr and In concentrations are as low as $\sim3$\% and $\sim6$\% respectively, difficult to be detected by standard characterization methods. A direct and unambiguous way to probe the relation between chemical inhomogeneities and low-field anomalies is to perform new measurements after an appreciable change of these inhomogeneities.\cite{picos}  In the case of underdoped LSCO, the strong $T_C$-dependence on the Sr concentration prevented us to obtain samples of the same nominal composition but with sharper transitions. However, in the case of the Pb-In alloys, we have checked that the low-field anomalies are significantly reduced in a more homogeneous Pb$_{92}$In$_8$ alloy, and they are completely absent in pure Pb. These results fully confirm our present experimental results and interpretations.

The results presented here do not definitively exclude a possible explanation in terms of some of the intrinsic mechanisms proposed by other authors for the enhanced fluctuation diamagnetism observed in different cuprate superconductors under low reduced temperatures and magnetic fields. \cite{prb_carretta,lascialfari02,lascialfari03,Ong1,Ovchinnikov,Sewer,demello,Anderson} However, the observation of similar anomalies in slightly inhomogeneous conventional (singlet s-wave pairing) BCS low-$T_C$ superconducting alloys at least weakens these proposals. Complementarily, the results presented in this paper provide an important check of our previous conclusions for the underdoped LSCO studied here:\cite{prl_pc_lasco} In all the $h-t$ region no too close to the superconducting transition, the measured precursor diamagnetism is explained at a quantitative level by the GGL approach, empirically extended to high reduced temperatures by introducing a \textquotedblleft total energy\textquotedblright cutoff. \cite{prl_pbin,epl_ybco,uncertainty,prb_pbin} 
These conclusions suggest that the precursor diamagnetism is not affected by the presence of a pseudogap in the normal state, in agreement with recent paraconductivity measurements in cuprates with different doping levels.\cite{Curras,Naqib} 
However, new measurements of $\Delta M(T,H)$ in these cuprates will be desirable to confirm the generality of these last conclusions. 
\newline
\newline
\begin{acknowledgments}
We acknowledge the financial support from the Spanish \textit{Ministerio de Educaci\'on y Ciencia and FEDER funds} under Grant No. MAT2004-04364, and from the \textit{Xunta de Galicia} under Grant No. PGIDIT04TMT206002PR. L. C. acknowledges financial support from Spanish \textit{Ministerio de Educaci\'on y Ciencia} through a FPU grant. F. S. acknowledges financial support from \textit{Uni\'on Fenosa} under grant 220-0085-2002.
\end{acknowledgments}

\pagebreak

\pagebreak

\begin{table}
\caption{\label{tab:table1}Main superconducting parameters for the samples studied. $\overline T_C$ and $\Delta T_C$ were obtained from low-field $\chi^{FC}(T)$ measurements (see Fig. \ref{FC} and inset of Fig. \ref{plomos}). In the case of Pb and the Pb-In alloys, the critical fields and $\kappa$ were obtained from $M(H) $measurements below $T_C$.\cite{prb_pbin,cm_Nb} In the case of LSCO, $H_{C2}(0)$ comes from the analysis of the superconducting fluctuations above $T_C$,\cite{prl_pc_lasco} and $H_{C1}(0)$ from $M(H)$ measurements below-$T_C$.\cite{prb_lasco} }
\begin{ruledtabular}
\begin{tabular}{lcccccc}
Sample & $\overline T_C$ & $\Delta T_C$ & $\mu_0H_{C1}(0)$ & $\mu_0H_{C2}(0)$ & $\mu_0H_C(0)$ & $\kappa$\\
  & (K) & (K) & (mT) & (T) & (T) &  \\
\hline
La$_{1.9}$Sr$_{0.1}$CuO$_4$ & 26.3 & 2.75 & 16 & 26.2 & \textemdash &$60$\footnotemark[1]\\
$\rm{Pb_{55}In_{45}}$ & 6.47 & 0.06 & 38\footnote{Obtained from the relation $H_{C1}(0)=H_{C2}(0)\ln\kappa/2\kappa^2$.} & 1.20 & \textemdash &5.1\\
Pb$_{92}$In$_8$ & 7.00 & 0.02 & 31\footnotemark[1] & 0.49 & \textemdash &2.1 \\
Pb & 7.17 & 0.01 & \textemdash & \textemdash &  0.14 & 0.3 \\
\end{tabular}
\end{ruledtabular}
\end{table}


\begin{references}
\bibitem{Lascialfari_comment}Although these authors indicate that $H_{\rm peak}$ seems to increase with temperature, at least for temperatures farther from $T_C$ than $\Delta T_C$ (a behavior against any explanation in terms of $T_C$ inhomogeneities), a closer inspection of their data indicate that for those temperatures the $M(H)$ curves do not present a diamagnetic peak at the indicated $H_{\rm peak}$, but just a change in the slope at a much smaller magnetic field. This last behavior can be easily explained with the $T_C$-inhomogeneities model.
\end{references}
\end{document}